\pdfoutput=1
\documentclass[a4paper,12pt]{article}
\usepackage{amsmath,amssymb,amsfonts,mathtools,amsthm}
\usepackage{hyperref}

\usepackage{graphicx}
\usepackage{comment}
\usepackage{placeins}
\usepackage{todonotes}
\usepackage{algorithmicx,algorithm,algpseudocode}
\usepackage{subcaption}
\usepackage{authblk}
\usepackage{tikz}
\usepackage{cite}

\newcommand{\ket}[1]{\ensuremath{|#1\rangle}}
\newcommand{\bra}[1]{\ensuremath{\langle#1|}}
\newcommand{\ketbra}[2]{\ensuremath{\ket{#1}\bra{#2}}}
\newcommand{\proj}[1]{\ensuremath{\ket{#1}\bra{#1}}}
\newcommand{\braket}[2]{\ensuremath{\langle{#1}|{#2}\rangle}}

\newcommand{\ii}{\mathrm{i}}

\newcommand{\Hl}{\mathcal H}

\newcommand{\R}{\mathbb R}

\newcommand{\ie}{\emph{i.e.\/}}

\newtheorem{theorem}{Theorem}

	\title{Impact of global and local interaction on quantum spatial search on chimera graph}
	\author[1,2]{Adam Glos\footnote{\url{aglos@iitis.pl}}}
	\author[1,3]{Tomasz Januszek}
	\affil[1]{Institute of Theoretical and Applied Informatics, Polish Academy of 
		Sciences, ul. Ba{\l}tycka 5, 44-100 Gliwice, Poland}
	\affil[2]{Institute of Informatics, Silesian University of Technology,
		ul. Akademicka 16, 44-100 Gliwice, Poland}
	\affil[3]{Institute of Mathematics, Silesian University of Technology, Kaszubska 23, 44-100 Gliwice, Poland}
	
\date{}
\begin{document}

\maketitle

\begin{abstract}
In the paper, we investigated the influence of local and global interaction on
the efficiency of continuous-time quantum spatial search. To do so, we analyzed
numerically chimera graph, which is defined as 2D grid with each node replaced
by complete bipartite graph. Our investigation provides a numerical evidence
that with a large number of local interactions the quantum spatial search is
optimal, contrary to the case with limited number of such interactions. The
result suggests that relatively large number of local interactions with the
marked vertex is necessary for optimal search, which in turn would imply that
poorly connected vertices are hard to be found.
\end{abstract}

\section{Introduction} \label{sec:introduction}

Quantum spatial search  is an example of a quantum algorithm outperforming any
classical one. Since the very first paper\cite{childs2004spatial}, many graphs
were shown to be efficiently searchable
\cite{childs2004spatial,glos2018vertices,chakraborty2016spatial}. Still, there
is no general simple condition on graph verifying if the continuous-time quantum
spatial search runs optimally--the known ones base on spectral properties of
graph matrices and not on the properties of graph topology
\cite{chakraborty2016spatial,chakraborty2018finding}. In fact, most of the
results contradict simple conditions like connectivity
\cite{meyer2015connectivity} or global symmetry \cite{janmark2014global}.

Furthermore, small effort has been made on a characteristics representing
real-world interactions. Such graphs have typically three features: they are
small-world, they have power-law degree distribution and clustering property.
While the radius of the marked vertex seems to have impact on the efficiency of
quantum spatial search \cite{philipp2016continuous}, there is not much effort
made on the last two graph characteristics. As for clustering property, simplex
of complete graphs was only analyzed \cite{wong2015diagrammatic}, however the
construction does not allow changing the cluster's size. 

In this paper we analyze how the ratio between local interactions within the
cluster and global interactions  between clusters influences the efficiency of
continuous-time quantum spatial search. To do so we choose the chimera graph
which represents the topology used on D-Wave quantum
computer~\cite{boixo2014evidence}. The chimera graph $\chi(a,b,c)$ is defined as
$a\times b$ grid graph with each node replaced by complete bipartite graphs
$K_{c,c}$. Such subgraph consists of $2c$ vertices and $c^2$ edges, which makes
it a dense subgraph, \ie{} a cluster. In contrast to bipartite graphs, the grid
graph is a sparse graph, which result in small number of interactions between
clusters.

In the paper we show that the bigger the cluster is, and by this the more
significant the local interactions within the cluster are, the faster the
quantum search is. The result suggests that large number of local interactions
is needed in order to make the quantum search efficient. Contrary, it may be
difficult to find poorly connected vertices.

The paper is organized as follows. In Sec.~\ref{sec:preliminaries} we explain
basic facts concerning the quantum spatial search and chimera graph. In
Sec.~\ref{sec:experiment} we describe our experiment and analyze our result. In
Sec.~\ref{sec:conclusion} we present general conclusions and discuss their
possible extensions.

\section{Theoretical preliminaries} \label{sec:preliminaries}

Let $G=(V,E)$ be an undirected graph with $n=|V|$ vertices. Let $\Hl$ be a
$n$-dimensional spaces spanned by an orthogonal basis $\{\ket v:v\in V\}$.
Furthermore, let $w\in V$ be a marked vertex. The continuous-time quantum
spatial search is defined by a Hamiltonian \cite{childs2004spatial}
\begin{equation}
H = -\gamma A_G - \ketbra{w}{w},
\end{equation}
where $A_G$ is an adjacency matrix of the graph, and $\gamma$ is a hopping rate that
needs to be derived for a graph. The hopping rate needs to be of order
$\Theta(1/\|A_G\|)$, where $\|A_G\|$ is spectral norm. Otherwise the $A_G$ or
oracle $\ketbra{w}{w}$ would play dominant role and the other Hamiltonian part
would be ignored. The evolution starts in the uniform superposition
$\ket{\psi_0} = \frac{1}{\sqrt{n}}\sum_{v\in V} \ket v$, which reflects the lack
of knowledge of the position of the marked vertex. It ends with a measurement in
computational basis. The success probability equals $p(t)=|\bra w \exp(-\ii t
H)\ket{\psi_0}|^2$, and the algorithm for its derivation is presented in
Alg.~\ref{alg:success_probability}

\begin{algorithm}[t]
	\caption{Derives the success probability of quantum spatial search}	\label{alg:success_probability}
	\begin{algorithmic}
		\State \textbf{Input}: $G$---graph, $w\in V(G)$---marked vertex, $\gamma$---hopping rate, $t$---evolution time
		\State \textbf{Output}: success probability $p(t)$
		\Function{succ\_probability}{$G$, $w$, $\gamma$, $t$}
		\State $H \leftarrow - (\gamma A_{G} + \proj{w})$ \Comment $A_G$ is adjacency matrix of $G$
		\State $\ket {\psi_0} \leftarrow \frac{1}{\sqrt{|V(G)|}}\sum_{v\in V(G)}\ket v$
		\State $p \leftarrow |\bra w \exp(-\ii t H) \ket{\psi_0}|^2 $
		\State \Return $p$
		\EndFunction
	\end{algorithmic}
	
\end{algorithm}

\begin{figure}
\centering
\begin{minipage}{0.45\textwidth}\centering
	\includegraphics{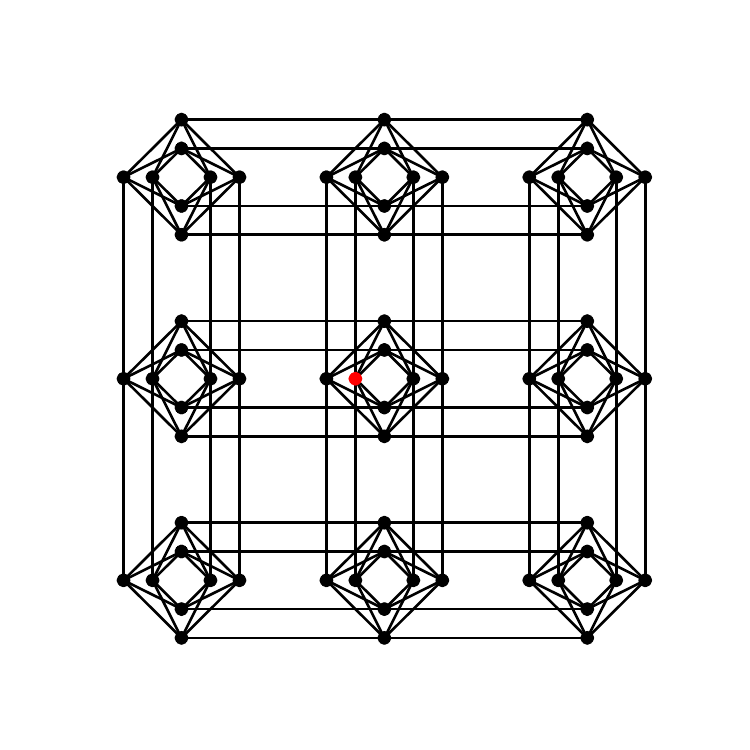}
\subcaption{Chimera graph $\chi(3,3,4)$}
\label{fig:chimera-3}
\end{minipage}
\qquad
\begin{minipage}{0.45\textwidth}\centering
	\includegraphics{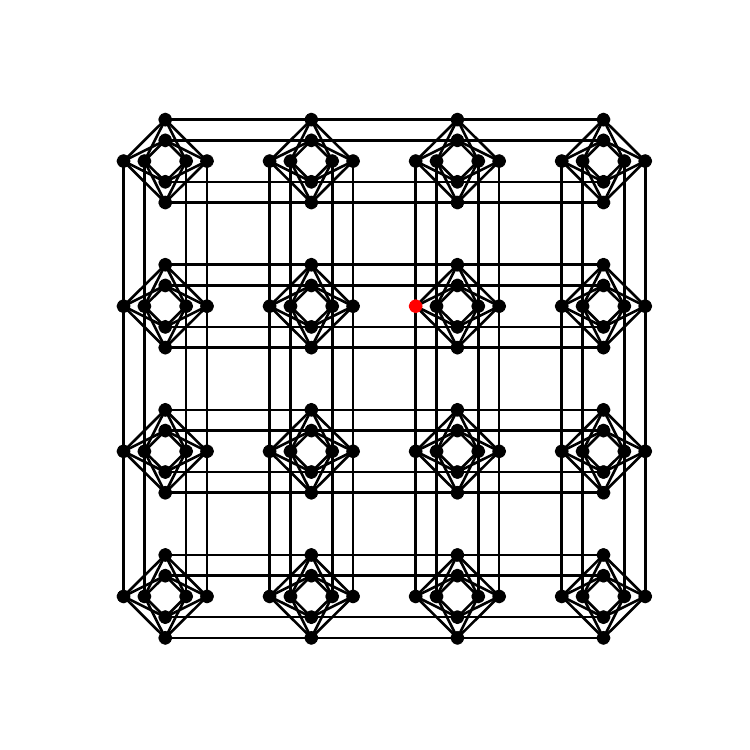}
	\subcaption{Chimera graph $\chi(4,4,4)$}
	\label{fig:chimera-4}
\end{minipage}

\caption{The chimera graphs $\chi(k,k,4)$ together with their marked elements (red dots).
	The marked element is always the element in the center of the chimera
	graph.}\label{fig:chimera-graphs}
\end{figure}

In the scope of this paper we will analyze the square chimera graph
$\chi(k,k,l)$. The graph was analyzed before in term of simple quantum
walk~\cite{xu2018quantum}. It consists of $n=2k^2l$ vertices and $\sim 2k^2l^2$
edges. The graph is constructed as $k\times k$ two-dimensional grid graph, in
which each node is replaced with complete bipartite graph $K_{l,l}$, see
Fig.~\ref{fig:chimera-graphs}. Note that bipartite graphs can be considered as
dense graphs, as they have $l^2$ edges with only $2l$ vertices. Contrary there
is only $4l$ edges connecting bipartite graphs, $l$ between the bipartite graph
and each bipartite graphs located at left, right, top and bottom. Because of
this we call the bipartite graphs clusters. Note that half of the  vertices of
bipartite graph are connected with clusters at top and bottom, and the rest $l$
is connected to left and right cluster. 

Note that searching on a 2D grid graph is a
hard task \cite{childs2004spatial}, contrary to the complete bipartite graph
case, were search is quantumly optimal \cite{novo2015systematic}. Since the
order of both graphs can be changed by the chimera graph parameters, they enable
the analysis of impact of local interaction coming from the `smaller' bipartite
graphs and global interactions coming from the grid graph.

In our analysis we will consider the case where the marked element is placed at
the center of the graph, see Fig.~\ref{fig:chimera-graphs}. Note that there
exists a very fast classical algorithm for finding such element, especially for
small bipartite subgraphs. However our aim is not to show that the complexity of
quantum spatial search is smaller comparing to any classical search, as it has
been done before for other graphs
\cite{chakraborty2016spatial,childs2004spatial}. Instead, we plan to analyze how
the change of the order of local interaction influences the quantum spatial
search efficiency, which makes our choice well-justified.

Three approaches are usually considered for measuring the efficiency of the
quantum spatial search. In the first we analyze the complexity of time $t$
required for obtaining state $\ket{\psi_t} \approx \ket{w}$. After such
evolution the measurement outputs $w$ with probability close to 1 \cite{chakraborty2016spatial}, or at least very high comparing to initial probability $1/n$~\cite{glos2018optimal,glos2018vertices,childs2004spatial}. In general
such approach may require more complicated tools \cite{wong2015diagrammatic},
which makes the numerical analysis intractable, or at least very difficult. Furthermore, maximizing the success probability may influence the complexity of evolution time, which in turn should be minimized. 

Instead of this multiple-criteria approach, in the scope of this paper we
minimize $t/p(t)$ function. It has simple interpretation as repeating the search until
the correct vertex is measured. Expectedly, for fixed $t$ we need $1/p(t)$
repetitions, which after including the time cost result in mentioned formula.
The third approach is using the amplitude amplification
\cite{brassard2002quantum}, which could be considered as quantum-like repetition
approach. Since last two approaches are very similar in their concept, and they
differ with $\sqrt{p(t)}$ factor, we will consider only the $t/p(t)$ approach in
the rest of the paper.

Note that $t/p(t)$ achieves global minimum at $t=0$. This unintuitive result comes
from neglecting the time cost of algorithm preparation and measurement. While in
the case of analytical analysis it can be ignored, the numerical calculation
converge to this uninformative result in most cases. In order to remove the minimum,
we change the cost function into
\begin{equation}
 \mathcal{T}(t) = \frac{t + t_{\rm penalty}}{p(t)},
\end{equation}
where $t_{\rm penalty}$ denotes the cost of preparation and measurement. This
cost function was already used in \cite{glos2018impact}. Typically $t_{\rm penalty} \sim c
\log(n)$ is sufficient for removing the  convergence  to $t=0$ in most cases.

Now we will show that theoretical results proposed in
\cite{chakraborty2016spatial,chakraborty2018finding} cannot be applied in the
context of chimera graphs. Let us recall the statement presented in these
references.
\begin{theorem}[\cite{chakraborty2016spatial}] \label{lem:the-lemma}
	Let $H_1$ be a Hamiltonian with eigenvalues $\lambda_1\geq\dots\geq\lambda_n$ satisfying $\lambda_1=1$ and $|\lambda_i|\leq \Delta <1$ for all $i>1$ with corresponding eigenvectors $\ket {v_1},\dots,\ket{v_n}$ and let $w$ denote a marked vertex. For an appropriate choice of $r\in[-\frac{\Delta}{1+\Delta},\frac{\Delta}{1-\Delta}]$, applying the Hamiltonian $(1+r)H_1 +\ketbra{w}{w}$ to the starting state $\ket{s}$ for time $t = \Theta(\sqrt n)$ results in the state $\ket f $ with $|\braket{w}{f}|\geq \sqrt{\frac{1-\Delta}{1+\Delta}}+o(1)$.
\end{theorem}
\begin{theorem}[\cite{chakraborty2018finding}]
	Let $H_1$ be a Hamiltonian with eigenvalues $\lambda_1=1>\lambda_2=1-\Delta\geq \dots\geq \lambda_n\geq 0$ such that $H_1\ket{v_i} = \lambda_i\ket{v_i}$. Let $H_{\rm oracle} = \ketbra{w}{w}$ with $w =\sum_{i}a_i\ket{v_i}$ and $|\braket{w}{v_1}|=|a_1|=\sqrt{\varepsilon}$. Let $r=\sum_{i\neq 1}\frac{|a_i|^2}{1-\lambda_i}$ and $\nu = \frac{r}{\sum_{i\neq 1}\frac{|a_i|^2}{(1-\lambda_i)^2}}$. Provided $\sqrt{\varepsilon} \ll r\Delta /\nu$, then evolving the state $\ket{v_1}$ under the Hamiltonian $rH_1+H_{\rm oracle}$ for time $t=\Theta \left( \frac{1}{\sqrt{\varepsilon}\nu}\right)$, results in a state $\ket{f}$ with $|\braket{w}{f}\approx \nu$.
\end{theorem}

\begin{figure}\centering
	\includegraphics[]{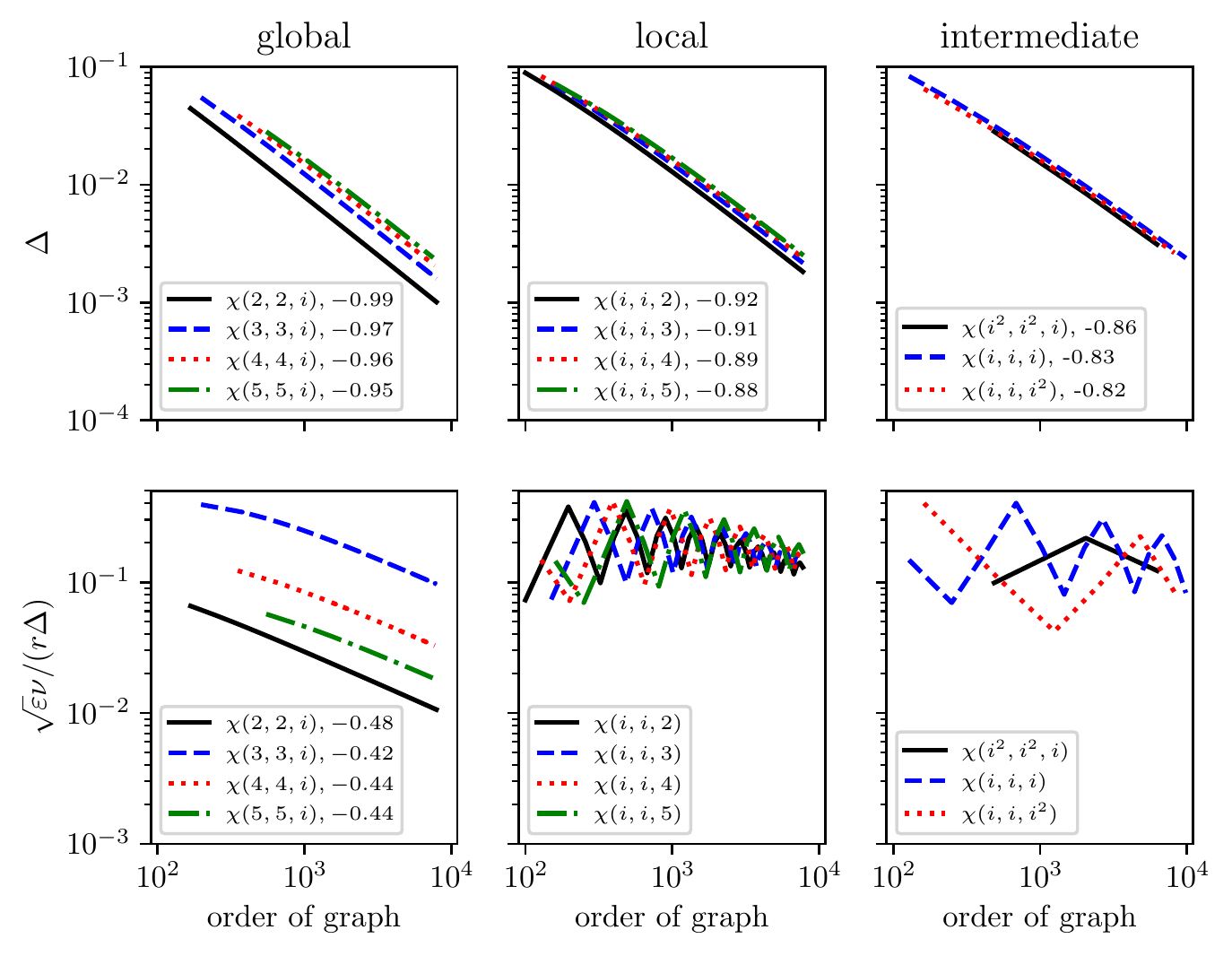}
	\caption{\label{fig:Chakraborty-conditions}The conditions for fast quantum spatial search from \cite{chakraborty2016spatial} and \cite{chakraborty2018finding}. The definitions of $\varepsilon$, $\nu$, $r$ and $\Delta$ can be found in Sec.~\ref{sec:preliminaries}. Note that $\Delta$ decreases for any kind of chimera sequences, while the second measure decreases only for global chimera graphs. The second value in the legend (if present) is the slope of the regression line.}
\end{figure}
\begin{figure}\centering
	\includegraphics[]{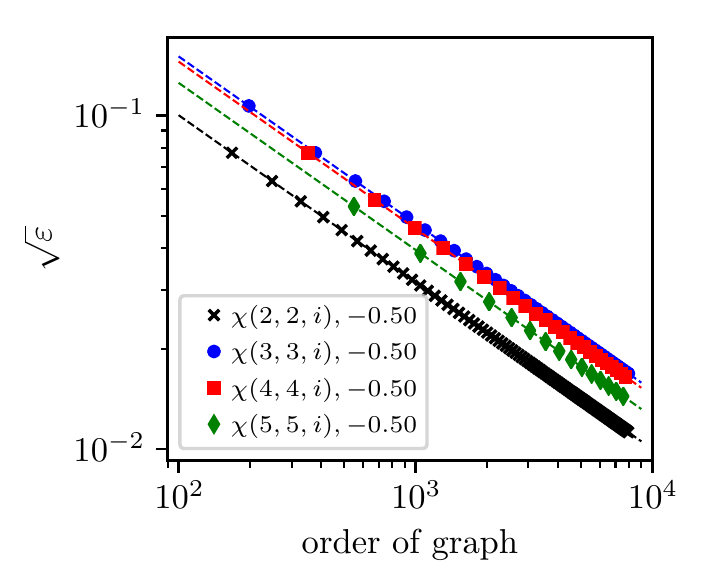}\includegraphics[]{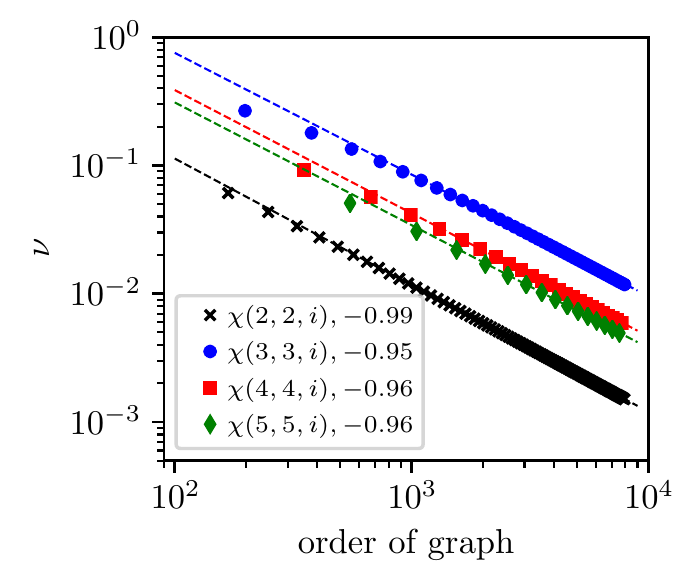}
	\caption{\label{fig:Chakraborty-eps-nu}The values of $\sqrt{\varepsilon}$ and $\nu$ for global chimera graph. Dashed lines are regression lines for the $\log(\varepsilon)= \alpha \log(n)+\beta$ model, where $n$ is the graph order, similarly for $\nu$. The second value in the legend is the slope of the regression line.}
\end{figure}

In Fig.~\ref{fig:Chakraborty-conditions} we present our numerical approach on the conditions verification for multiple chimera graph classes. In particular, we have investigated `local' chimera graphs, where the size of the cluster is constant, `global', where the grid size is constant, and `intermediate', where both cluster and grid part grows. We have calculated the $\Delta,\sqrt\varepsilon, \nu$ and $r$ for `centralized' Hamiltonian $H_1=aA+b$, where $a,b\in \R$ were chosen in such a way that $|\lambda_2|=|\lambda_n|$, and by this $\Delta$ is maximized. Such approach was already used before \cite{glos2018vertices,glos2018optimal}. We can see that for none of the graphs the $\Delta= \Theta(1)$ holds, which is a required condition for Theorem~\ref{lem:the-lemma} to hold. 

The condition  $\sqrt{\varepsilon} \ll r\Delta /\nu$ holds only for global chimera graphs, provided in the local and intermediate cases the result will remain constant. However based on results presented in Fig.~\ref{fig:Chakraborty-eps-nu}, it implies that the efficiency of the algorithm is $t/p(t) = \Theta(1/\sqrt{\varepsilon}\nu^3) \approx \Theta(n^{3.35})$, which is much worse than the results coming from our numerical simulations, which will be shown in next section.

\section{Numerical analysis} \label{sec:experiment}

\subsection{Data generation}
\begin{figure}
\begin{algorithm}[H]
	\caption{Derives the upperbound for evolution time for chimera graph $\chi(k,k,l)$. \textsc{Marked} returns marked element as described in Sec.~\ref{sec:preliminaries}}	\label{alg:time_upperbound}
	\begin{algorithmic}
		\State \textbf{Input}: $k$---grid parameter, $l$---bipartite  subgraph parameter
		\State \textbf{Output}:
		\Function{time\_upperbound}{$k$, $l$}
		\State $G \leftarrow \chi(k,k,l)$
		\State $w\leftarrow \textsc{marked}(G)$
		\State $t_{\rm init} \leftarrow 1$ 
		\State $p_{\rm init} \leftarrow \textsc{succ\_probability}(G, w, \frac{1}{l+1}, t_{\rm init})$
		\State $t_{\max}' \leftarrow t_{\rm init}/p_{\rm init}$
		\State ${\rm result} \leftarrow t_{\max}'$
		\For {$t=\frac{1}{10}t_{\max}';\frac{2}{10}t_{\max}';\dots; t_{\max}' $}
		\State ${\rm result } \leftarrow \min ({\rm result , t/\textsc{succ\_probability}(G, w, \frac{1}{l+1}, t)})$  
		\EndFor
		\State \Return result
		\EndFunction
	\end{algorithmic}
\end{algorithm}

\begin{algorithm}[H]
	\caption{Derives the list of $(\gamma_{\rm opt}, t_{\rm opt})$, resulting in local optimum $\mathcal{T}_n$. The function \textsc{NelderMead}$(f,\ p,\ \rm bounds)$ find the locally optimal point $(\gamma_{\rm out},t_{\rm out})$ for function $f$ starting from initial point $p$ and restricting the optimization to bounds. For purpose of our paper we have used the default implementation from \texttt{Optim.jl} \cite{mogensen2018optim}. \textsc{Marked} returns marked element as described in Sec.~\ref{sec:preliminaries}}\label{alg:qss_optimization}
	\begin{algorithmic}
		\State \textbf{Input:} $k$---grid parameter, $l$---bipartite subgraph parameter
		\State \textbf{Output:} list of the $(\gamma_{\rm opt}, t_{\rm opt})$ resulting in local optimum $\mathcal{T}_n=t_{\rm opt}/p(t_{\rm opt})$
		\State \texttt{result} $\leftarrow \emptyset$
		\Function{qss\_optimization}{$k$, $l$}
		\State $t_{\rm bound} \leftarrow \textsc{time\_upperbound}(k,l)$
		\State $w\leftarrow \textsc{marked}(\chi(k,k,l))$
		\For {$\gamma_0=\frac{1}{20\|A_{\chi(k,k,l)}\|},\frac{2}{20\|A_{\chi(k,k,l)}\|},\ldots,\frac{2}{\|A_{\chi(k,k,l)}\|}$}
		\For {$t_0=0,\frac{1}{15}t_{\rm bound},\ldots, t_{\rm bound}$}
		\State $f(\gamma, t) =(t+\log(n))/\textsc{succ\_probability}(\chi(k,k,l),w,\gamma,t) $
		\State $(\gamma_{\rm out},t_{\rm out}) \leftarrow \textsc{NelderMead}(f, (\gamma_0,t_0), [[0,1],[0,t_{\rm bound}]])$
		\State append $(\gamma_{\rm out},t_{\rm out})$ to \texttt{result}
		\EndFor
		\EndFor
		\State \Return \texttt{result}
		\EndFunction
	\end{algorithmic}
\end{algorithm}
\end{figure}

\begin{figure}[h]\centering
	\includegraphics{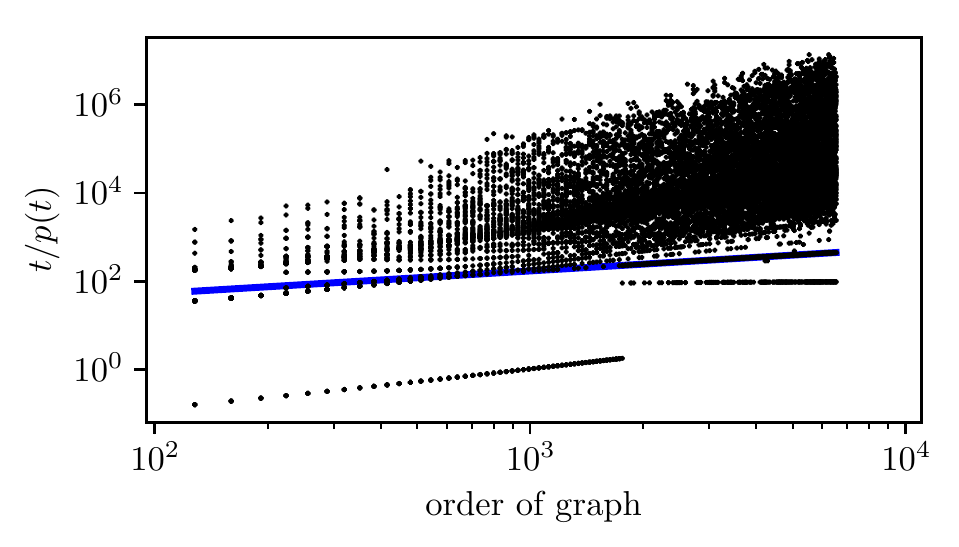}
	\caption{Example of raw, unfiltered data and fitted linear regression. The linear regression represented by blue line is made on the minimum $t/p(t)$ values for each graph order, ignoring the point for which $t<\log(n)/n$ or $p(t)<\log(n)/n$. The slope $\alpha$ of the regression line represents the algorithm's complexity $\Theta(n^\alpha)$.}\label{fig:unfiltered-data}
\end{figure}
In the preliminary numerical calculation we optimize the function
\begin{equation}
\mathcal{T}_n(\gamma, t) = \frac{t + 2\log(n)}{p(\gamma,t)}.
\end{equation}
Note that theoretically $\gamma,t\in \R_{\geq 0}$, nevertheless the problem can be
easily converted into optimization problem on the two-dimensional interval. First note, that
$\gamma\in[0,1]$. Secondly, let $t_{\rm opt}$ denotes optimal measurement time.
Then
\begin{equation}
\mathcal{T}_n(\gamma, t) \geq\mathcal{T}_n(\gamma_{\rm opt}, t_{\rm opt})=\frac{t_{\rm opt} + t_{\rm penalty}}{p(\gamma_{\rm opt},t_{\rm opt})} \geq t_{\rm opt} +t_{\rm penalty}\geq  t_{\rm opt}.
\end{equation}
The inequality implies that the search problem can be reduced to $t\in
[0,\mathcal{T}_n(\gamma, t)]$, where $t$ and $\gamma$ are chosen arbitrarily.
Hence the optimization problem on unbounded quadrant can be reduced to optimization
on convex set $(\gamma,t)\in[0,1]\times [0,\mathcal{T}_n(\gamma, t)]$.

Let us define the algorithm determining the optimal point $(\gamma_{\rm opt},
t_{\rm opt})$ minimizing  $\mathcal{T}_n(\gamma, t)$. We start by determining
the upper-bound on the evolution time, which is presented in
Alg.~\ref{alg:time_upperbound}. We start by making initial upperbound $t'_{\rm
	bound}=T_n(1/(l+1),1)$. Then we improve the bound by choosing the minimum over
expected times with different time evolution values. Note that theoretically
both $\gamma$ and $t$ can be chosen arbitrarily, however our preliminary experiments
suggested that $\frac{1}{l+1}$ is good initial value. Note that the
choice is consistent with $l\to \infty$ limit requirement, since
$\|A_{\chi(k,k,l)}\| \approx \Delta\leq l+2$, where $\Delta$ is maximum degree.
We end up the upperbound derivation by determining the minimum of the $\mathcal
T_n (\frac{1}{l+1}, t)$ for $t=0.1\,t'_{\rm bound},0.2\,t'_{\rm
	bound},\ldots,0.9\,t'_{\rm bound}$. By this we obtained the final upperbound $t_{\rm bound}$.

In the next step we have granulated the $[0,1]\times [0,t_{\rm bound}]$ and
using Nelder-Mead method we have determined the local optima points
$(\gamma_{\rm out},t_{\rm out})$. For optimization we have used the default
implementation \textsc{NelderMead} from Julia module \texttt{Optim.jl}
\cite{mogensen2018optim}. Exemplary unfiltered data are presented in
Fig.~\ref{fig:unfiltered-data}. We can observe that despite the penalty there
are still points with small $t/p(t)$ value. Hence we have removed every
$(\gamma, t)$ tuple, for which $t<\log(n)/n$ or $p(t)<\log(n)/n$. From the rest
we took minimum for each order of the graph. We claim that these pairs
represent a configuration close to optimal for quantum spatial search.

\begin{figure}[!th]
	\centering
	\includegraphics[scale=0.85]{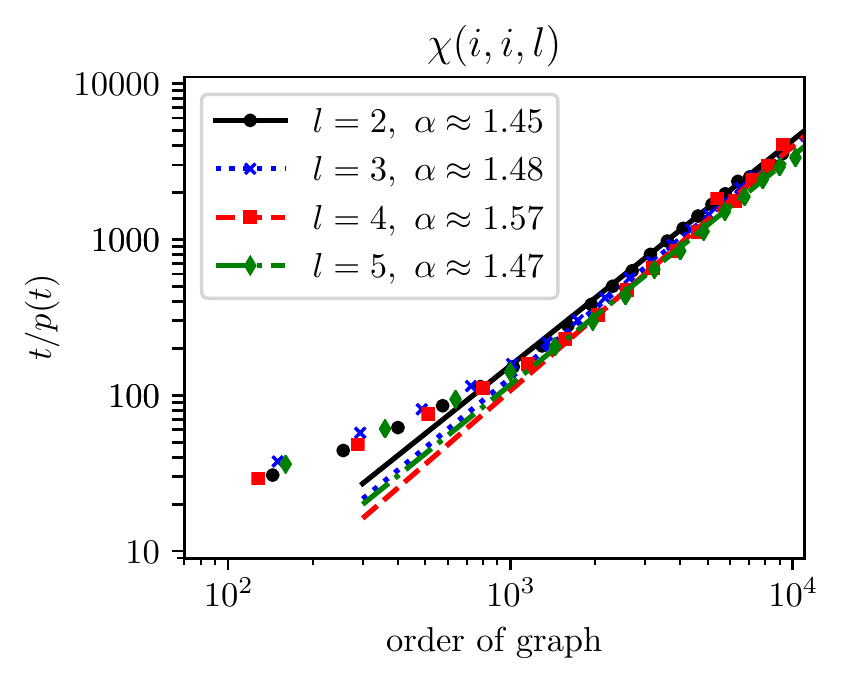}
	\includegraphics[scale=0.85]{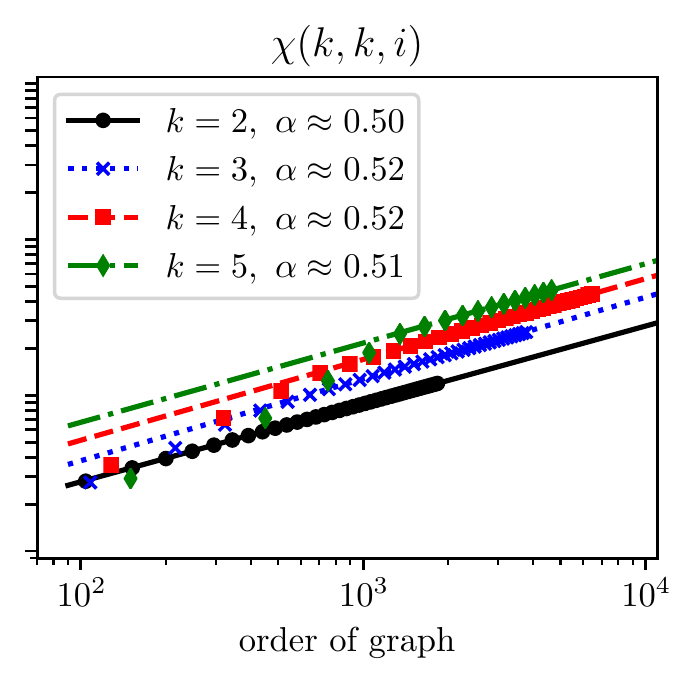}
	
	\caption{The optimal points and the linear regression for the  $\chi(i, i, l)$ case and the $\chi(k,k,i)$ the. Note that in the second case the quantum spatial search is optimal, while on the left it is worse than classical random guess.}\label{fig:regression}
	
	\includegraphics{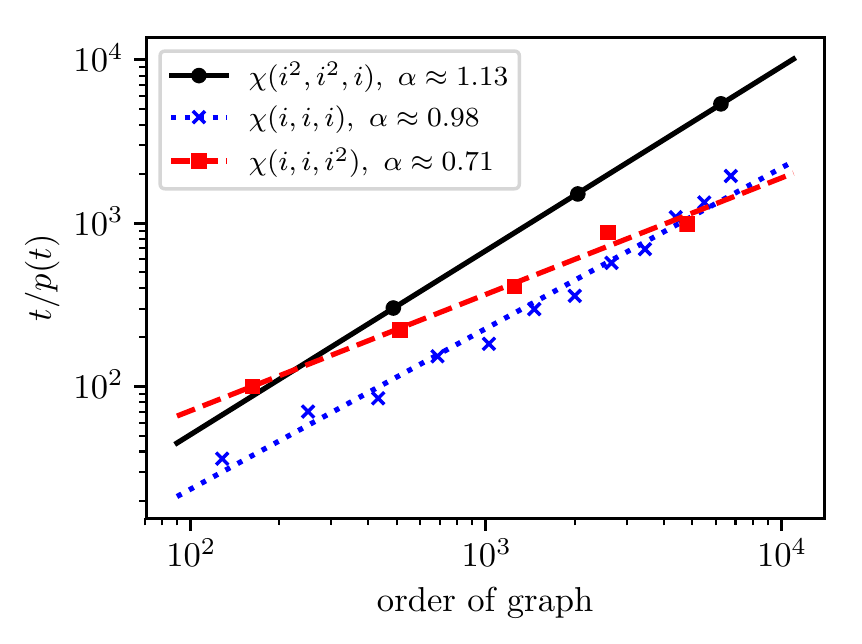}
	\caption{Intermediate interaction case. Note the slope $\alpha$ depends on the ratio between local and global interaction case} \label{fig:global}
\end{figure}

\subsection{Result analysis}
Suppose the complexity of the graph can be approximated by $\Theta(n^\alpha)$. It means that 
\begin{equation}
\mathcal{T} = c n^{\alpha}(1+o(1)),
\end{equation}
which can be transformed into
\begin{equation}
\log(\mathcal{T}) = \log(c(1+o(1))) + \alpha \log (n).
\end{equation}
Hence slope of line regression of $\log(\mathcal{T})$ vs $\log(n)$ provides an approximation of the complexity.

We have generated the data for chimera graphs $\chi(i,i,l)$ for $l=2,3,4,5$
called `local', and for chimera graphs $\chi(k,k,i)$ with $k=2,3,4,5$ called
`global', both with consecutive values of $i$. For each case we have determined
the $\alpha$ parameter. The result are presented in Fig.~\ref{fig:regression}.

We can observe that in the first case the complexity is roughly
$\Theta(n^{1.5})$, which is worse than random guess. This shows that if the
local interactions are small, the negative impact of the grid prevents fast
search. In the case of large number of local interaction coming from
bipartite graph topology, the search is at least close to optimal, and influences
only the constant next to the leading term of the evolution.

To continue our approach we have considered intermediate interaction case
$\chi(i^2,i^2,i)$, $\chi(i,i,i)$ and $\chi(i,i,i^2)$. Note that these example
are transition case between global and local chimera graphs, hence we would
expect that the quantum spatial search efficiency should change smoothly between
worst $\Theta(n^{1.5})$ and optimal $\Theta(\sqrt{n})$ complexities. Based on
the numerical results presented in Fig.~\ref{fig:global} we can see that it is
indeed the case. The quantum search on $\chi(i,i,i^2)$ graph, which is
the closest to local chimera graphs of all intermediate graphs considered, is
not quantumly optimal anymore. However at the same corresponding complexity is
better compared to other intermediate cases. Accordingly, $\chi(i,i,i)$ provides
worse complexity, and $\chi(i^2,i^2,i)$ has the worst complexities among all of
analyzed intermediate cases. These results suggests that the complexity of the
fraction of cluster and graph size has direct impact on algorithm efficiency.

Note that $\Theta(n)$ is in fact the worst possible complexity, as it refers to
measuring at a very small time. This may suggest that our result
$\Theta(n^{1.5})$ implies badly designed numerical analysis. However, as we
mentioned in Sec.~\ref{sec:preliminaries}, we have dropped the results with
small $t$ and $p(t)$ values which would result in linear complexity.
Furthermore we would like to emphasize that our goal was not to design optimal
quantum spatial search on chimera graph, but to analyze the effect of cluster
size change on the algorithm efficiency. Finally based on the results from
Fig.~\ref{fig:regression} and \ref{fig:global} we see that even dropping time
complexity to $\Theta(n)$ in all these cases doesn't change our conclusions,
because we are interested in complexity change instead of the actual order.

\section{Conclusion and discussion}\label{sec:conclusion}

In the paper we analyzed the continuous-time quantum spatial search on chimera
graph, which represents the D-Wave computer topology. We showed that the bigger
the clusters are, and thus there is more local interactions represented by
bipartite graph cells are, the better the efficiency of the algorithm is. We
claim that efficient search of marked element may require many local
interactions.

One way to extend the results is to provide analytical derivation of our
numerical simulation. Furthermore, it would be interesting to analyze graphs
with more complex topology. In particular, stochastic block model is a natural
extension.

\paragraph{Acknowledgments} Tomasz Januszek acknowledges the support by the
Polish National Science Center under the Project Number 2014/15/B/ST6/05204.
Adam Glos acknowledges the support by the Polish National Science Center under
the Project Number 2016/22/E/ST6/00062. The authors would like to thank Piotr
Gawron for revising the manuscript and discussion.

\bibliographystyle{ieeetr}
\bibliography{chimera}

\end{document}